\documentclass[preprint]{ptephy_v1}

\preprintnumber{} 

\begin{document}

\title{Search for Majorana neutrinos}


\author{Itaru Shimizu}
\affil{Research Center for Neutrino Science, Tohoku University, Sendai 980-8578, Japan \email{shimizu@awa.tohoku.ac.jp}}

\begin{abstract}
Whether there exist elementary particles having Majorana nature is one the fundamental open question that has persisted since the 1930s. The only practical experiments to test the Majorana nature of neutrinos is the search for neutrinoless double-beta decay, which has been a major challenge for nuclear and particle physicists. In the 2000's, a number of experiments using advanced technologies have been planned, and some of which have already achieved significant improvements in the search sensitivity. In this article, the current status of the neutrinoless double-beta decay searches are summarized, reviewing the progress of KamLAND-Zen, which recorded the world best sensitivity in the effective Majorana neutrino mass limit.
\end{abstract}

\subjectindex{C43, D02}

\maketitle

\section{Introduction}

Neutrinos are elementary particles that have no electric and color charges, and interact weakly with other particles. According to standard cosmology, neutrinos are the most abundant fermions in our universe. Understanding the properties of neutrinos is of great importance for particle physics and cosmology, however, their experimental study is not easy because of their rare interactions with matter. After long-standing efforts, observations of neutrino oscillation have revealed that neutrinos have tiny masses, the first discovery of physics beyond the Standard Model (SM). Since neutrino masses are much smaller than those of other fermions, the origin of neutrino masses must be something special. An uncharged neutrino particle could be its own antiparticle, the so-called Majorana neutrino, proposed by Ettore Majorana in 1937. This hypothesis is important to build a theoretical mechanism for tiny neutrino masses: right-handed heavy neutrinos with Majorana masses on the GUT scale naturally lead to light neutrino masses (See-Saw mechanism), however, so far there is no experimental evidence to support the theory. In addition, the $CP$ violating decay of the heavy neutrinos in the early Universe can also explain the matter dominance in the universe (Leptogenesis). The Majorana neutrino is a key piece of particle physics and cosmology, and the determination of its relevant parameters, the masses and $CP$ phases of the Majorana neutrino, is of critical importance. In this article, I review the recent status of the experimental search for the Majorana neutrino.

\section{Majorana neutrino}

\begin{figure}[t]
\begin{center}
\includegraphics[width=0.8\columnwidth]{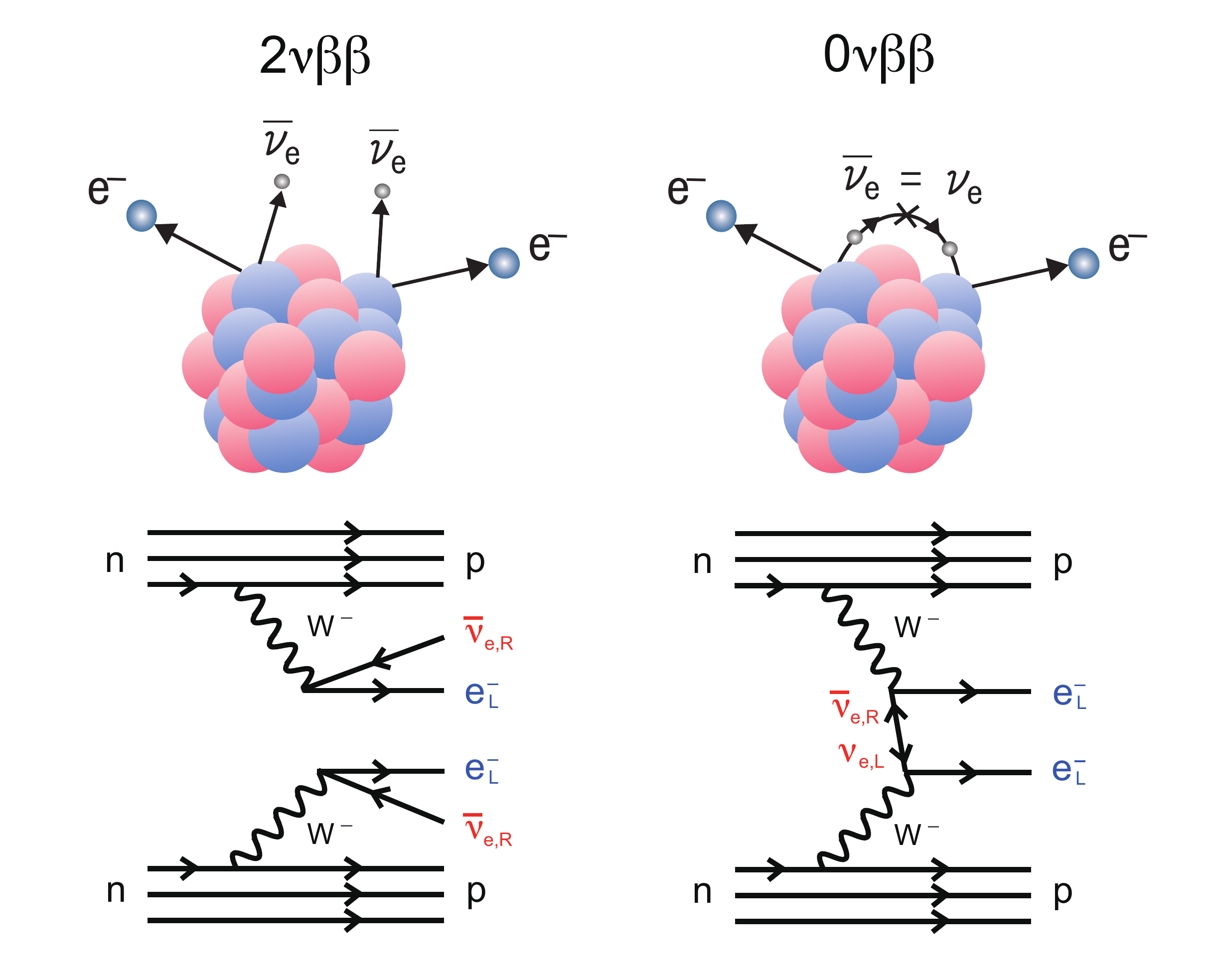}
\end{center}
\vspace{-0.5cm}
\caption{Decay scheme and Feynman diagrams for $2\nu\beta\beta$ (left) and $0\nu\beta\beta$ (right).}
\label{figure:decay_scheme}
\end{figure}

Neutrinos were first introduced by W.~Pauli in 1930 to reconcile an anomalies in the conservation of energy and spin in nuclear beta-decay. Experimental observation of neutrinos has been difficult for decades because they interact with other particles only through weak interactions. The double-beta ($\beta\beta$) decay, equivalent to two simultaneous $\beta$ decays, was first proposed by M.~Goeppert-Mayer in 1935~\cite{Mayer1935}, soon after the quantitative theory of $\beta$-ray emission by E.~Fermi. The double-beta decay is observable, if one decay is suppressed by energy levels or spin states, but the two simultaneous decays are not. In 1937, E.~Majorana suggested the possibility of an uncharged neutrino particle being its own antiparticle. Based on this hypothesis, in 1939, W.H.~Furry proposed the idea of double-beta decay with no neutrino emission at all, now called neutrinoless double-beta decay ($0\nu\beta\beta$)~\cite{Furry1939}, 
\begin{equation}
(A, Z) \rightarrow (A, Z\!+\!2) + 2 e^{-}, \label{eq:0nbb}
\end{equation}
emitting only two electrons, explicitly violating lepton number by two (Fig.~\ref{figure:decay_scheme}). $0\nu\beta\beta$ decay can be realized when two electron anti-neutrinos annihilate inside nucleus, while such process is forbidden in the SM, and can be an excellent probe of the Majorana neutrino.

On the other hand, the SM-allowed second-order nuclear decay is two neutrino double-beta decay ($2\nu\beta\beta$), 
\begin{equation}
(A, Z) \rightarrow (A, Z\!+\!2) + 2 e^{-} + 2 \overline{\nu}_{e}, \label{eq:2nbb}
\end{equation}
emitting two electrons and two electron anti-neutrinos (Fig.~\ref{figure:decay_scheme}). Since $0\nu\beta\beta$ decays do not emit anti-neutrinos and have a monoenergetic energy peak at $Q$-value, they can be experimentally distinguished from $0\nu\beta\beta$ by observing the total energy of the two electrons. The first experimental search was conducted by E.L.~Fireman in 1948 using Geiger counters and 25\,g of enriched $^{124}$Sn~\cite{Fireman1948}. Subsequent experiments used a variety of particle detectors, such as Geiger, proportional, and scintillation counters. The initial $0\nu\beta\beta$ search experiments often found false positive signals, later disproved by other experiments. After the discovery of maximal parity violation in weak interactions in 1957, the predicted decay rate for $0\nu\beta\beta$ was highly suppressed than that for $2\nu\beta\beta$, so the search for Majorana neutrinos via $0\nu\beta\beta$ was recognized for a time to be a challenging task.

The discovery of neutrino oscillation is a hopeful sign for the $0\nu\beta\beta$ search, as  $0\nu\beta\beta$ may be mediated by an exchange of the light Majorana neutrinos, as shown in Fig.~\ref{figure:decay_scheme}. In the case of a massless Majorana neutrino, right-handed $\overline{\nu}_{e}$ emitted from a neutron has the wrong helicity ($h = +1$) for absorption on another neutron, however, neutrinos should have finite masses to cause neutrino oscillation, so its helicity-flip can occur ($h = -1$) with a probability depending on their masses. In the model of light Majorana neutrino exchange, the $0\nu\beta\beta$ decay rate (inverse half-life) is 
\begin{equation}
( T_{1/2}^{0\nu} )^{-1} = G_{0\nu} \left| M_{0\nu} \right|^{2} \left<m_{\beta\beta}\right>^{2}, \label{eq:0nbbrate}
\end{equation}
where $G_{0\nu}$ is the phase-space factor, $M_{0\nu}$ the nuclear matrix element (NME), $\left<m_{\beta\beta}\right>$ the effective Majorana neutrino mass. $G_{0\nu}$ can be accurately calculated based on the wave function of the electron in the Coulomb field of the nucleus. On the other hand, since the evaluation of the NME requires dealing with complex nuclear structure, so the theoretical estimate of NME highly depends on the many-body approach used in the calculation. Considering the dependence on calculation method, the value of $\left| M_{0\nu} \right|$ will have an uncertainty by a factor of 2 to 3. 

As indicated in Eq.~(\ref{eq:0nbbrate}), the decay rate increases with the square of the effective Majorana neutrino mass
\begin{equation}
\left<m_{\beta\beta}\right> = \left| \sum_{i} U_{ei}^{2}m_{\nu_{i}} \right|, 
\end{equation}
where $U_{ei}$ is the $3 \times 3$ unitary neutrino mixing matrix, and $m_{\nu_{i}}$ are the masses of three neutrinos ($\nu_{1}, \nu_{2}, \nu_{3}$). It is important to note that this sum includes complex $CP$ phases in $U_{ei}$ (one Dirac and two Majorana $CP$ phases), so cancellations can occur. Measurements of neutrino oscillation have determined the 3 mixing angles ($\theta_{12}, \theta_{23}, \theta_{13}$) and 2 square mass differences ($0 < \Delta m^{2}_{21} \ll |\Delta m^{2}_{32}|$, where $\Delta m^{2}_{ij} = m_{\nu_{i}}^{2} - m_{\nu_{j}}^{2}$), while the ordering of three neutrino masses (sign of $\Delta m^{2}_{32}$) is still uncertain. Thus, there are two possibilities of the mass ordering in the three neutrinos, the normal ordering (NO) with one heavier neutrino relative to the other two ($m_{\nu_{1}} < m_{\nu_{2}} \ll m_{\nu_{3}}$), or the inverted ordering (IO) with two heavier neutrinos ($m_{\nu_{3}} \ll m_{\nu_{1}} < m_{\nu_{2}}$). If the lightest mass is zero, the predicted range of $\left<m_{\beta\beta}\right>$ based on best-fit values of neutrino oscillation parameters~\cite{NuFIT2020} is 1$-$4\,meV and 19$-$48\,meV for NO and IO (Fig.~\ref{figure:effective_mass}), respectively, where the uncertainty in $\left<m_{\beta\beta}\right>$ is due to the Majorana $CP$ phases. Thus, the experimental constraints on $\left<m_{\beta\beta}\right>$ provide information on the absolute neutrino mass scale and ordering. 

In general, order 1\,ton of $\beta\beta$ isotopes are required to obtain a sensitivity of  $\left<m_{\beta\beta}\right>$  sensitivity down to 10$-$20\,meV. The sensitivity that an experiment can reach depends on the background level achieved in the experiment, and the NME and phase space factor of the isotope selected. Most of the near-future experiments aim at probing the IO region. To probe down to a few meV corresponding to the NO region, experiments would need to increase their exposures by about two orders of magnitude using order 100\,ton of $\beta\beta$ isotopes (Fig.~\ref{figure:effective_mass}). 

The absolute neutrino mass can also be probed by other approaches that do not assume the Majorana nature of the neutrino. Precise measurements of the electron energy spectrum in $\beta$ decay are sensitive to the effective neutrino mass
\begin{equation}
\left<m_{\beta}\right> = \sqrt{\sum_{i} \left| U_{ei} \right|^{2}m_{\nu_{i}}^{2}}. 
\end{equation}
The KATRIN experiment using tritium $\beta$ decay reported the latest result, $\left<m_{\beta}\right> < 0.8\,{\rm eV}$ at 90\% confidence level (C.L.)~\cite{Aker2022}. In the future, KATRIN aims to reach a sensitivity of $0.2\,{\rm eV}$~\cite{Aker2022}, which covers the predicted $\left<m_{\beta}\right>$ of IO. Furthermore, since massive neutrinos strongly influence the evolution of the large-scale structure of the Universe, cosmological observations can place very strong constraints on the sum of the three neutrino masses
\begin{equation}
\Sigma m_{\nu} = \sum_{i} m_{\nu_{i}}. 
\end{equation}
The Planck collaboration reported a stringent upper limit of $\Sigma m_{\nu} < 0.12\,{\rm eV}$ at 95\% C.L.~\cite{Aghanim2020}. A positive value of $\Sigma m_{\nu}$ from future observations would provide another milestone. If a positive $0\nu\beta\beta$ signal is found at the rate expected from other observations, it could prove the scenario of the light Majorana neutrino exchange. Even if negative, the contradiction with other observations would disprove the standard scenario, and could have a significant impact on particle physics and cosmology.

\section{$0\nu\beta\beta$ search experiments}

The basic idea of the $0\nu\beta\beta$ search is to observe two emitted electrons and find a monoenergetic peak in the energy spectrum of their sum. In order to achieve high sensitivity, various ideas were considered, utilizing new detector technologies and low background techniques developed in neutrino observations, $\beta\beta$ decay and dark matter searches. 

The main background sources are $2\nu\beta\beta$, environmental radioactivities, such as $^{238}$U,  $^{232}$Th, and cosmogenic isotopes produced by muon spallation. The background of solar neutrinos should be taken into account if the target isotope concentration is small. The main features for significant background reduction are as follows, 
\begin{enumerate}
\item Target nucleus increase per volume (isotope enrichment)
\item Good energy resolution (narrow energy window for $0\nu\beta\beta$ signal)
\item Thick shielding for external radiation (underground detector with active shield)
\item Particle identification (multisite energy deposits in $\gamma$ background)
\item Sequential decay tagging (background rejection with space and time correlation)
\item Daughter nucleus tagging ($0\nu\beta\beta$ signal identification)
\end{enumerate}

The status of leading $0\nu\beta\beta$ experiments in the world are introduced below.
\begin{itemize}
\item GERDA\\
In this experiment, a semiconductor detector technique using enriched $^{76}$Ge was developed. The detector consists of 44\,kg of high-purity germanium (HPGe) and is deployed into a  64\,m$^{3}$ cryostat containing ultra-pure liquid argon (LAr). It has a good energy resolution (FWHM) of 2.9\,keV at the $Q$-value (2039\,keV). The final GERDA result in 2020 provides a limit of $\left<m_{\beta\beta}\right> < 79-180\,{\rm meV}$ at 90\% C.L.~\cite{Agostini2020}. A future 1 ton detector in the LEGEND experiment will aim for a sensitivity down to $9-19\,{\rm meV}$~\cite{Adgrall2021}.
\item CUORE\\
The detector consists of 988 ultra-cold TeO$_{2}$ bolometers with natural tellurium for a total mass of 742\,kg (206\,kg of $^{130}$Te). The NTD thermistor converts the thermal pulse into a resistance variation, realizing a FWHM energy resolution of 7.8\,keV at the $Q$-value (2528\,keV). The CUORE result in 2022 provides a limit of $\left<m_{\beta\beta}\right> < 90-305\,{\rm meV}$ at 90\% C.L.~\cite{Adams2022}. The CUPID experiment can use the same cryogenic infrastructure as CUORE, for scintillating bolometers in replacing TeO$_{2}$ with Li$_{2}$MoO$_{4}$ enriched in the isotope of $^{100}$Mo, and will aim for a sensitivity down to $10-17\,{\rm meV}$~\cite{Alfonso2022}.
\item EXO\\
In this experiment, a liquid xenon time projection chamber (TPC), containing $\sim$175\,kg of xenon enriched to 80.6\% in $^{136}$Xe (EXO-200) was developed. The liquid xenon (LXe) TPC is capable of simultaneously reading ionization and scintillation. The event position is obtained from the time delay between the prompt light and the delayed charge signals, and also multisite energy deposits can be identified. The final EXO-200 result provides a limit of $\left<m_{\beta\beta}\right> < 94-286\,{\rm meV}$ at 90\% C.L.~\cite{Anton2019}. Cosmogenically produced $^{137}$Xe was a problematic background. Future plans to use 5 tons of enriched xenon in the nEXO experiment aim to have a sensitivity down to $6-18\,{\rm meV}$, assuming a deeper experimental location at SNOLAB~\cite{Albert2018}. The possibility of tagging Ba daughters is also being explored.
\item KamLAND-Zen\\
KamLAND has developed a xenon-loaded liquid scintillator detector (KamLAND-Zen), which has great advantages in the scalability of $\beta\beta$ isotope amount. Details of the detector design and results are described in the next section.
\end{itemize}
There are a number of $\beta\beta$ experimental projects utilizing new detector technologies and low background techniques, which are summarized in \cite{Dolinski2019}. Calorimeter experiments, {\it i.e.} $\beta\beta$ source and detector are the same, are semiconductors (\mbox{GERDA}, \mbox{MAJORANA}, \mbox{LEGEND}), bolometers (\mbox{CUORE}, \mbox{CUPID}, \mbox{AMoRE}), liquid TPC (\mbox{EXO}), loaded liquid scintillator (\mbox{KamLAND-Zen}, \mbox{SNO+}, ZICOS), inorganic scintillator (\mbox{CANDLES}), and high-pressure gas TPC (\mbox{NEXT}, \mbox{AXEL}, \mbox{PandaX-III}) experiments. On the other hand, external source experiments (NEMO) use thin sources which allow electrons to escape without significant energy loss. Although it is not easy to increase the source mass as in calorimeter experiments, external source experiments provide excellent angular correlation and individual energy measurements of the two emitted electrons. In the future, this feature will be important for distinguishing between the different underlying mechanisms of $0\nu\beta\beta$.

In this article, I focus on the results of the $0\nu\beta\beta$ search in KamLAND-Zen, which most stringently limits the Majorana neutrino mass.

\section{KamLAND-Zen}

\begin{figure}[t]
\centering\includegraphics[width=6.5in]{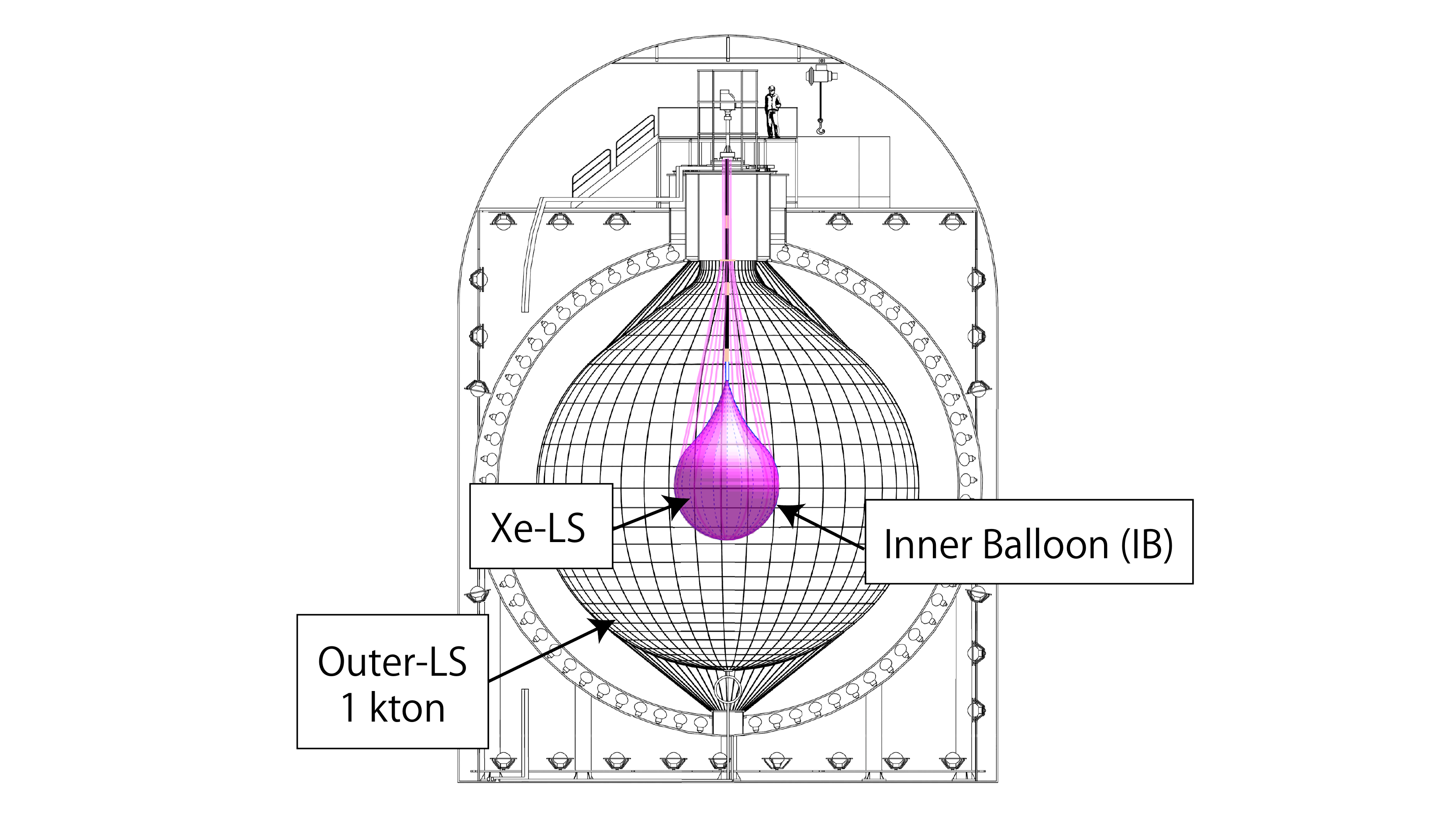}
\caption{Schematic diagram of the KamLAND-Zen detector.}
\label{figure:detector}
\end{figure}

\begin{figure}[t]
\centering\includegraphics[width=5.5in]{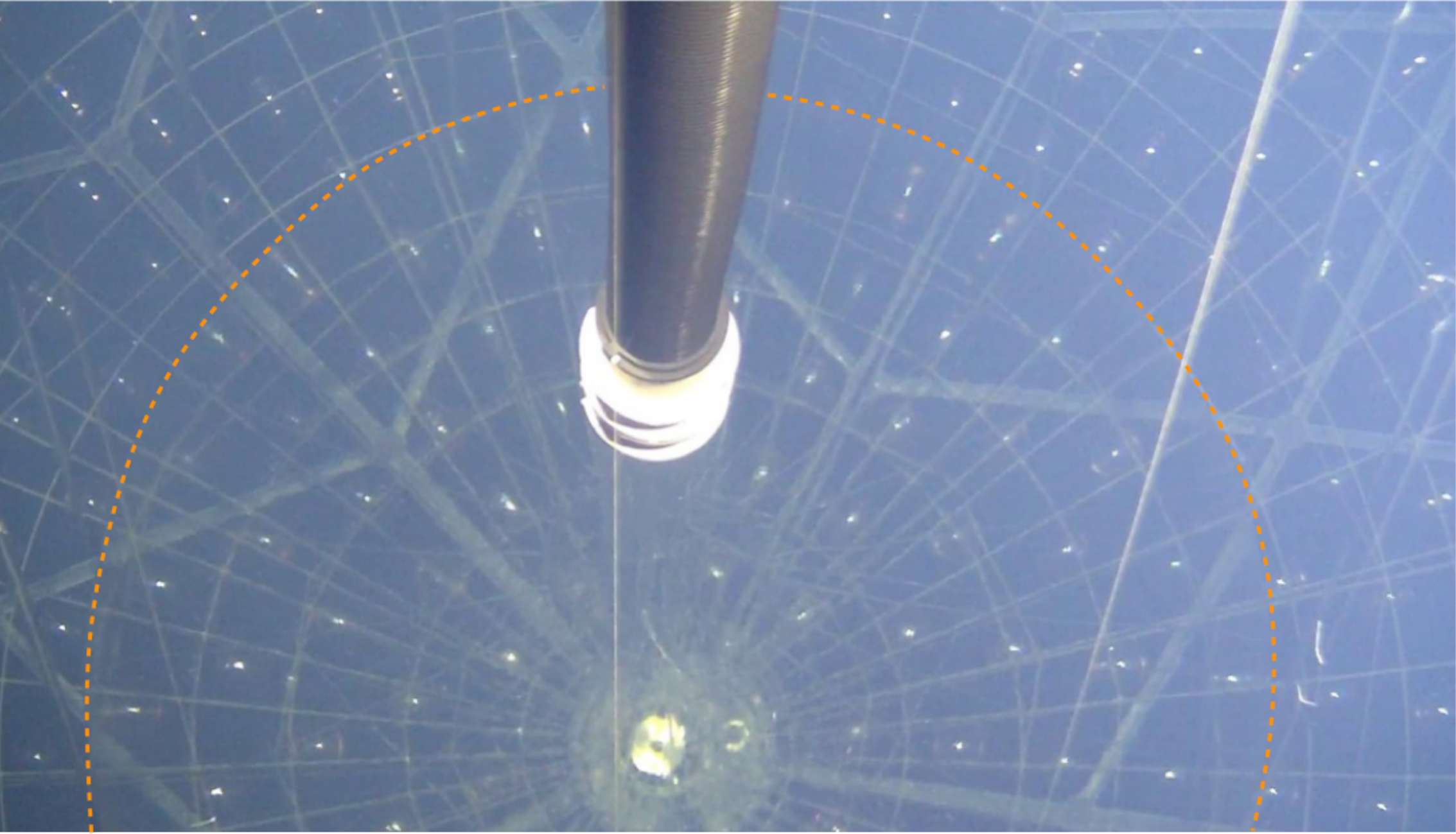}
\caption{Inside view of the KamLAND-Zen detector. The balloon was filled with the xenon-loaded liquid scintillator (Xe-LS). The dashed line shows the outline of the balloon shape.}
\label{figure:detector_inside}
\end{figure}

The KamLAND detector is a 1\,kton liquid scintillator detector located 1,000 m underground at the Kamioka mine in Japan, as illustrated in Fig.~\ref{figure:detector}. Scintillation light is viewed by 1,325 17-inch and 554 20-inch photomultiplier tubes (PMTs) mounted on an 18-m-diameter spherical stainless-steel tank. Taking advantage of its low background feature, KamLAND has achieved the world's first observation of reactor anti-neutrino oscillations~\cite{Eguchi2003,Abe2008} and the detection of geo-neutrinos~\cite{Araki2005b,Gando2011b}. This low background feature also allows for high sensitivity  $0\nu\beta\beta$ search with the addition of several detector components. Liquid scintillator (LS) can be loaded with $\beta\beta$ isotopes of $^{136}$Xe, which is best suited for use in KamLAND for the following reasons: (i) Isotopic enrichment of $^{136}$Xe gas by centrifugation is possible. (ii) Xenon gas is chemically stable and dissolves about 3\,wt\% in the LS. (iii) Slow $2\nu\beta\beta$ decay requires modest energy resolution. (iv) The well-known 2.614\,MeV $\gamma$ from $^{208}$Tl is not a serious background for $^{136}$Xe $0\nu\beta\beta$ ($Q$-value is 2.458\,MeV) because the coincident $\beta/\gamma$ from $^{208}$Tl are detected in homogeneous active detectors. Aiming at the $0\nu\beta\beta$ search in the IO region, 800\,kg of enriched xenon (91\% $^{136}$Xe) was prepared. The xenon was purified by distillation and refined with a heated zirconium getter. The water-dropped shape balloon was constructed by fixing 24 gores with a specially developed heat welding method. The folded ballon was sunk into the KamLAND detector through a narrow hole at the top, and inflated with the LS without xenon. Finally, the LS was replaced with the xenon-loaded LS (Xe-LS), as shown in Fig.~\ref{figure:detector_inside}.

The first data-taking of KamLAND-Zen began in 2011~\cite{Gando2012a}. The $\beta\beta$ decay source, 320\,kg of xenon was installed into a 3.08-m-diameter balloon. The first search showed event excess in the $0\nu\beta\beta$ signal region, which was identified as the background from $^{110m}$Ag decays, based on the energy spectrum and event rate variation~\cite{Gando2013a}. The $^{134}$Cs/$^{137}$Cs ratio revealed that the Xe-LS is contaminated by the Fukushima fallout. Since the sensitivity of the $0\nu\beta\beta$ search was limited by the background, $^{110m}$Ag was removed by Xe-LS purification~\cite{Shimizu2019}. First, xenon was extracted from the detector by LS circulation to confirm the remaining $^{110m}$Ag, and then the estimated background rate was cross-checked by On/Off (Xe-LS/Xe-depleted-LS) measurement. After the purification, 380\,kg of xenon was installed in 2013. The $0\nu\beta\beta$ search was continued until 2015, then most of the quasidegenerate neutrino mass region was surveyed~\cite{Gando2016}. To reach the IO region, the KamLAND-Zen detector was upgraded to larger Xe-LS volume (3.80-m-diameter balloon) containing 745\,kg xenon, corresponding to a twofold increase, and data-taking began in 2019. This search has provided a first test of the Majorana nature of neutrinos in the IO region.

I briefly discuss the latest result of the $0\nu\beta\beta$ search, which combines all KamLAND-Zen data up to May 2021~\cite{Abe2023}. The $2\nu\beta\beta$ decays are observed with the highest statistics, and the measured $2\nu\beta\beta$ decay half-life of $^{136}$Xe is $T_{1/2}^{2\nu} = 2.23 \pm 0.03({\rm stat}) \pm 0.07({\rm syst}) \times 10^{21}$\,year~\cite{Gando2019}. Due to the limited energy resolution, the resolution tail of the $2\nu\beta\beta$ events is one of the major background sources, as discussed below. The data with 745\,kg xenon contributes most significantly to the $0\nu\beta\beta$ search sensitivity. In this phase, the background due to radioactive impurities is dominated by $^{238}$U and $^{232}$Th daughter decays in the Xe-LS and balloon, and $^{110m}$Ag decays are absent. $^{214}$Bi decays ($^{238}$U daughter) are effectively removed by delayed coincidence tagging in the LS and are almost negligible, however, the untagged contributions on the balloon introduce backgrounds in the outer region within Xe-LS volume. In this phase, this contamination was reduced to approximately one-tenth of that in the previous phase~\cite{Gando2016} because the balloon was fabricated more cleanly. In the inner region, the background due to cosmogenic spallation products is the largest: cosmic-ray muons penetrating the Xe-LS can cause energetic reactions that produce spatially uniform isotopes from carbon and xenon spallation. The products with life-times greater than $O(100)$\,s, denoted as ``long-lived products'', are attributed to the decay of heavy isotopes produced by xenon spallation. Xenon spallation can be characterized by detecting multiple neutrons. Based on a likelihood-based discriminant, $(42.0 \pm 8.8)$\% of xenon spallation backgrounds are removed. The total live time is $523.4$\,days.

\begin{figure}
\vspace{0.1cm}
\begin{center}
\includegraphics[width=0.8\columnwidth]{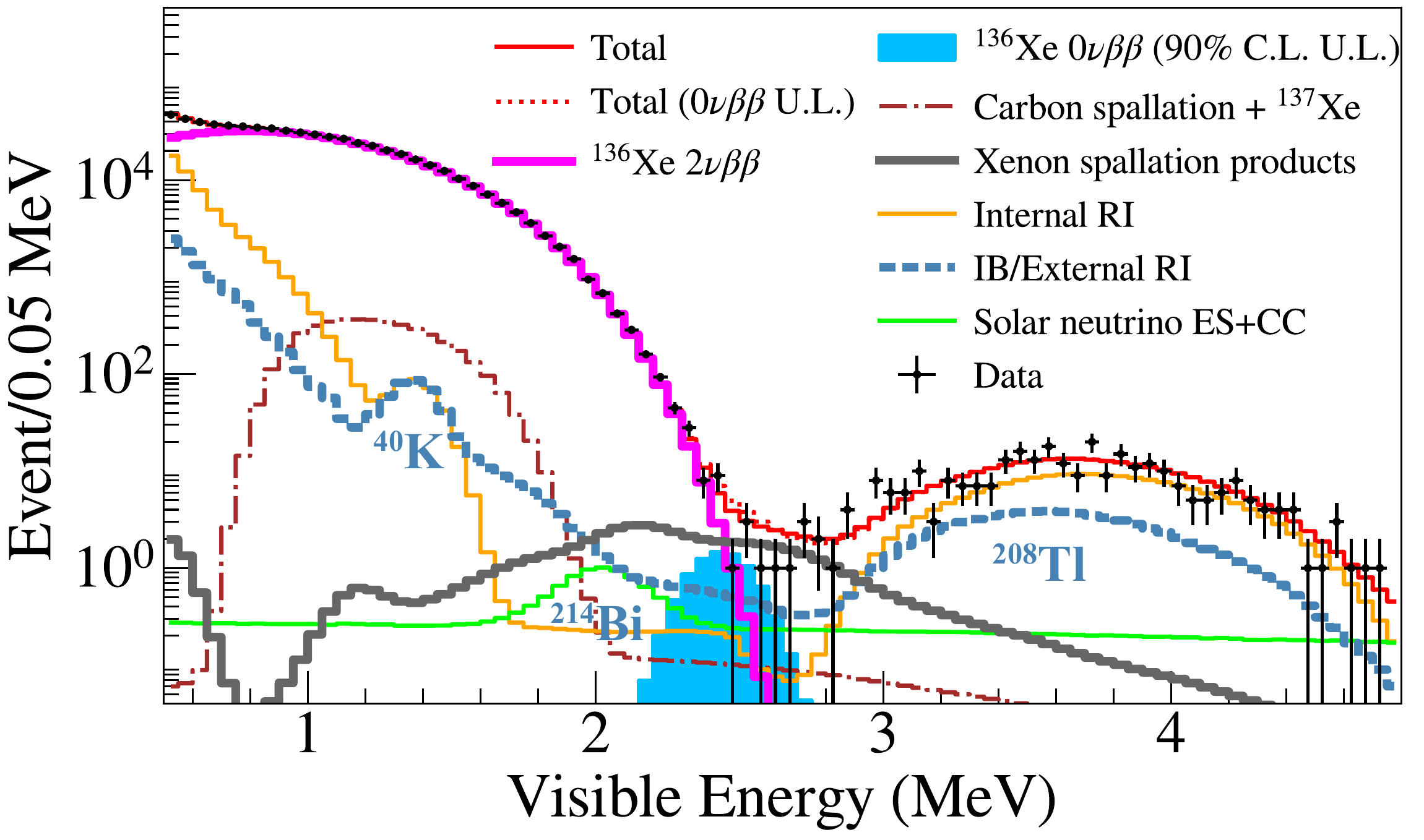}
\end{center}
\caption{Energy spectra of selected $\beta\beta$ candidates (singles data) within a 1.57-m-radius spherical volume drawn together with best-fit backgrounds, the $2\nu\beta\beta$ decay spectrum, and the 90\% C.L. upper limit for $0\nu\beta\beta$ decay. The primary backgrounds in the $0\nu\beta\beta$ region are $2\nu\beta\beta$, xenon spallation, and radioactive impurities (RI) mainly at the inner balloon (IB).}
\label{figure:energy}
\end{figure}

\begin{figure}[t]
\begin{center}
\includegraphics[width=0.8\columnwidth]{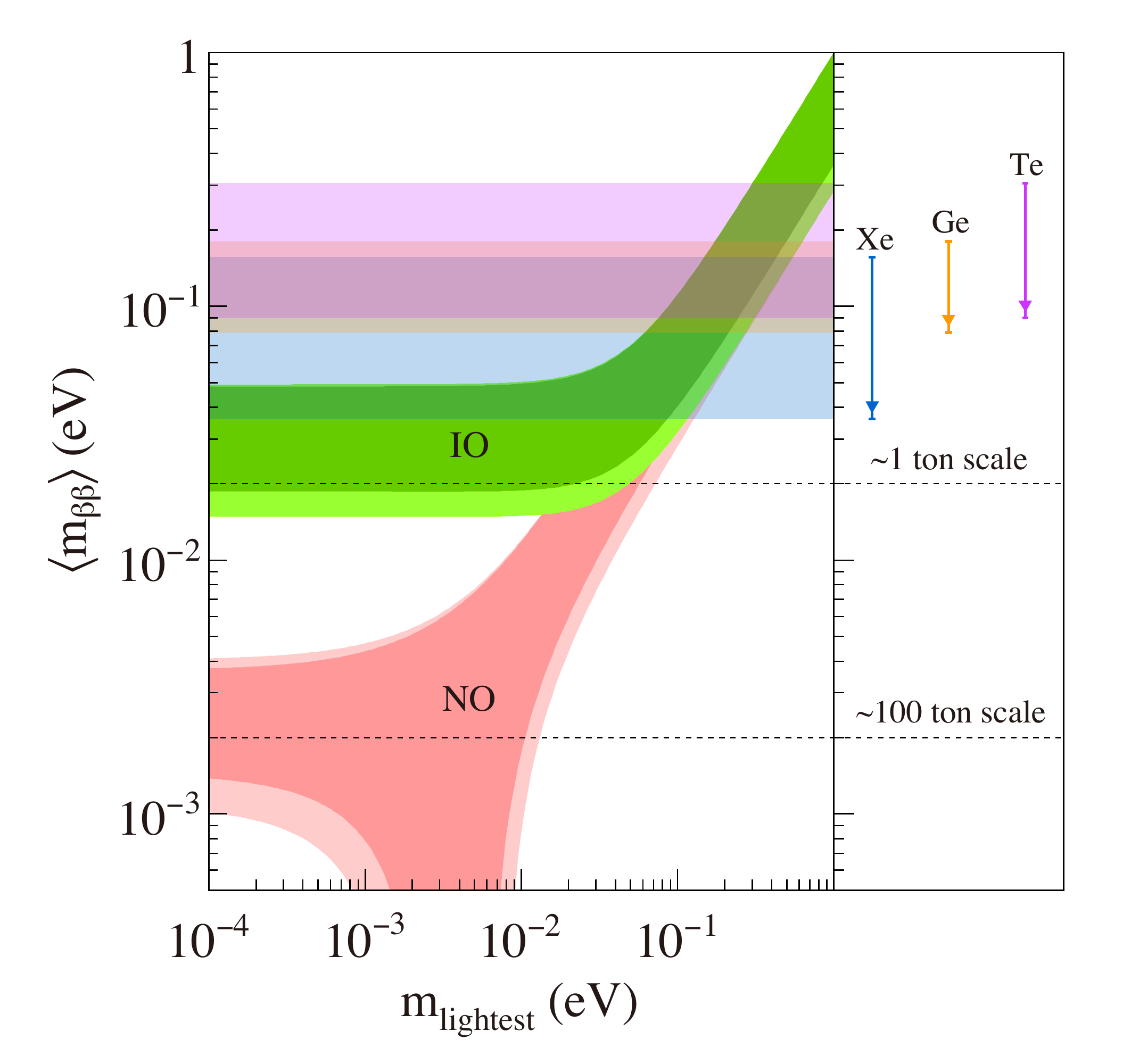}
\end{center}
\caption{Effective Majorana neutrino mass $\left<m_{\beta\beta}\right>$ as a function of the lightest neutrino mass $m_{\rm lightest}$. The dark shaded regions are predictions based on best-fit values of neutrino oscillation parameters for the normal mass ordering (NO) and the inverted mass ordering (IO), and the light shaded regions indicate the $3\sigma$ ranges calculated from oscillation parameter uncertainties~\cite{DellOro2014,NuFIT2020}. The horizontal bands indicate 90\% C.L. upper limits on $\left<m_{\beta\beta}\right>$ with $^{136}$Xe from \mbox{KamLAND-Zen}~\cite{Abe2023}, $^{76}$Ge~\cite{Agostini2020}, and $^{130}$Te~\cite{Adams2022}. The horizontal dashed lines indicate the required mass scale of $\beta\beta$ source to achieve deep exploration for each of the IO and NO regions.}
\vspace{-0.3cm}
\label{figure:effective_mass}
\end{figure}

Figure~\ref{figure:energy} shows the energy spectrum observed in KamLAND-Zen after removing the xenon spallation backgrounds. While there was no event excess from the $0\nu\beta\beta$ signal, the spectral fit provides the most stringent limit on the $^{136}$Xe $0\nu\beta\beta$ decay half-life of $T_{1/2}^{0\nu} > 2.3 \times 10^{26}$\,year at 90\% C.L., combining the analysis of the previous data-sets. The corresponding 90\% C.L. upper limit on the effective Majorana neutrino mass $\left<m_{\beta\beta}\right>$ is in the range $36 \text{ -- } 156$\,meV based on Eq.~(\ref{eq:0nbbrate}) using the phase space factor calculation from~\cite{Kotila2012,Stoica2013} and commonly used nuclear matrix element estimates~\cite{PhysRevLett.111.142501,PhysRevC.91.024316,Rodriguez2010,Deppisch2020,PhysRevC.91.034304,PhysRevC.101.044315,Horoi2015,Menendez2009,PhysRevC.102.044303,PhysRevC.91.024613,PhysRevC.87.045501,PhysRevC.87.064302,PhysRevC.97.045503} assuming the axial coupling constant $g_{A} \simeq 1.27$. Figure~\ref{figure:effective_mass} shows the prediction of $\left<m_{\beta\beta}\right>$ from neutrino oscillation parameters~\cite{DellOro2014,NuFIT2020} as a function of the lightest neutrino mass $m_{\rm lightest}$, together with the experimental limits from the $0\nu\beta\beta$ decay searches in $^{136}$Xe~\cite{Abe2023}, $^{76}$Ge~\cite{Agostini2020}, and $^{130}$Te~\cite{Adams2022}. The widths of the IO and NO bands reflect the uncertainties mainly due to the Majorana $CP$ phases, which cause cancellations of neutrino masses in $\left<m_{\beta\beta}\right>$. If no cancellations occur, the predicted values of $\left<m_{\beta\beta}\right>$ with $m_{\rm lightest} = 0$ are 4\,meV and 48\,meV for NO and IO, respectively. The search in $^{136}$Xe begins to test the IO band below 50\,meV, and realizes the partial exclusion of some theoretical models~\cite{Harigaya2012,Asaka2020,Asai2020}, that estimate $\left<m_{\beta\beta}\right>$ based on predictions of the $CP$ phases.

The sensitivity of the $0\nu\beta\beta$ search in \mbox{KamLAND-Zen} is limited by the primary background from xenon spallation products and $2\nu\beta\beta$ decay, as shown in Fig.~\ref{figure:energy}. In the current detector, multiple neutrons just after xenon spallation by muons are not fully detected due to a deadtime in electronics. In order to improve the removal efficiency of the xenon spallation background, state-of-the-art electronics that maximizes neutron detection efficiency is planned to be installed. Removal of xenon spallation background by particle identification is also a practical method. Most of the xenon spallation backgrounds are due to the radioactive decays which emit not only $e^{-}$ or $e^{+}$ ($\beta$ decay) but also coincident de-excitation $\gamma$'s or $e^{+}$-$e^{-}$ pair annihilation $\gamma$'s. Thus, background events from xenon spallation products have multisite energy deposits in the Xe-LS spreading over $O(10\,{\rm cm})$ distances due to $\gamma$-ray diffusion, while the $0\nu\beta\beta$ events are localized in smaller $<1\,{\rm cm}$ distances. Such dispersion of scintillation location can be statistically evaluated by using the photon hit time of the PMTs. To remove the xenon spallation backgrounds, an Integrated Spatiotemporal Deep Neural Network, referred to as KamNet~\cite{Li2023}, was developed in KamLAND-Zen. The simulation study using KamNet showed the ability to improve the $0\nu\beta\beta$ search sensitivity by the background removal.

\begin{figure}[t]
\centering\includegraphics[width=5.5in]{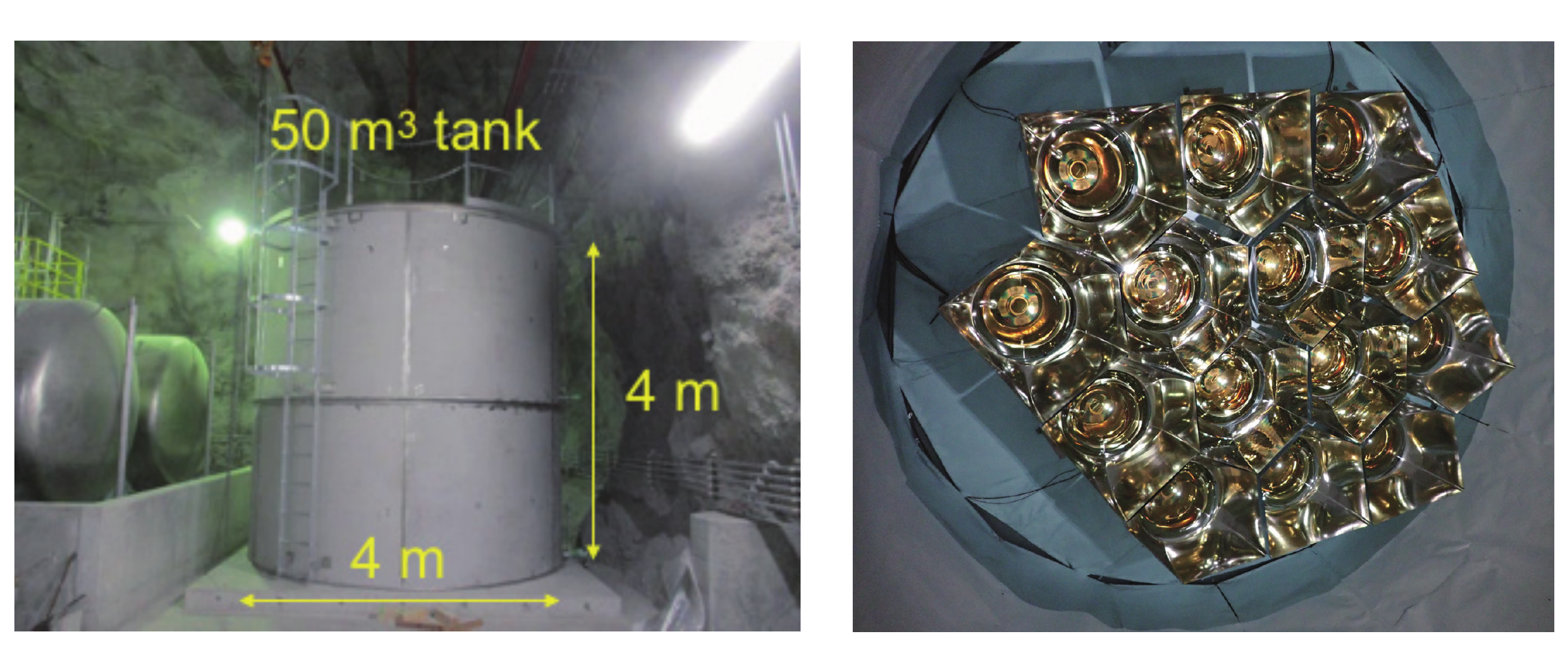}
\caption{KamLAND2-Zen prototype detector: (left) 50\,${\rm m}^{3}$ tank and (right) high quantum efficiency (HQE) PMTs with light collective mirrors in the tank.}
\label{figure:prototype}
\end{figure}

In the future, the KamLAND detector is planned to be upgraded to improve the $0\nu\beta\beta$ search sensitivity, denoted as \mbox{KamLAND2-Zen}. The discrimination between $0\nu\beta\beta$ and $2\nu\beta\beta$ can be improved by the upgraded detector with better energy resolution. The energy resolution ($\sigma$) improves from 4.0\% to $<$2.5\% at the $Q$-value of $^{136}$Xe $0\nu\beta\beta$ decay. This will be achieved by employing light collective mirrors ($>$1.8$\times$ light yield), new brighter LS (1.4$\times$ light yield), and high quantum efficiency (HQE) PMTs (1.9$\times$ light yield). To maximize light collection, new polygonal-shape mirrors, that do not have dead space in conventional circle-shape mirrors, were newly developed. To demonstrate this high detector performance, the \mbox{KamLAND2-Zen} prototype, which consists of a 50\,${\rm m}^{3}$ stainless tank, 14 HQE PMTs and light collective mirrors, new brighter LS, new state-of-the-art electronics, has been constructed at the Kamioka mine (Fig.~\ref{figure:prototype}). The data-taking in the prototype will begin soon. In addition, a scintillation balloon film made of Polyethylene naphthalate (PEN) has been also developed to remove the backgrounds from the balloon film by $\alpha$-tagging method. Also, research is underway to effectively reduce $\gamma$-emission backgrounds using a scintillation imaging system to identify multi-site energy deposits. To reduce solar neutrino background, a double concentration of $^{136}$Xe loading into the LS with pressurized xenon is also being investigated. The expected sensitivity in \mbox{KamLAND2-Zen} is $\left<m_{\beta\beta}\right> \sim$20\,meV, allowing a nearly full survey of the IO region, and the possibility of finding the $0\nu\beta\beta$ signal will be greatly enhanced.

\section{Conclusion}
Majorana neutrino, proposed more than 80 years ago, is still a major research topic in particle physics. In this article, I have described the current status and prospects of the $0\nu\beta\beta$ decay search. In particle physics and cosmology, the Majorana neutrino is the key to explaining the light neutrino masses and the matter dominance of the universe. In addition, $0\nu\beta\beta$ decay allows us to investigate the absolute neutrino mass, complementing searches in beta decay measurements and cosmological observations. Reflecting the importance of the $0\nu\beta\beta$ search, a number of experimental projects using modern technology are planned. The result in KamLAND-Zen, using an ultra-low background detector and an unprecedented 745\,kg of enriched xenon, has provided the most stringent limit on the Majorana neutrino mass, and the high sensitivity search is continuing. Although there was no experimental evidence of $0\nu\beta\beta$ decay, much progress has been made in the improved searches. Given the rapid progress in recent years, it would not be surprising if $0\nu\beta\beta$ decay is found in the near future.

\section*{Acknowledgment}

The author gratefully acknowledges the members of the KamLAND-Zen collaboration, providing information about their research activities and recent progress.

\let\doi\relax


\begin{thebibliography}{10}

\bibitem{Mayer1935}
M.~Goeppert-Mayer,
\newblock Phys. Rev. {\bf 48}, 512 (1935).

\bibitem{Furry1939}
W.~H. Furry,
\newblock Physical Review {\bf 56}, 1184 (1939).

\bibitem{Fireman1948}
E.~L. Fireman,
\newblock Phys. Rev. {\bf 74}, 1238 (1948).

\bibitem{NuFIT2020}
Nufit 5.0, available at http://www.nu-fit.org (2020).

\bibitem{Aker2022}
M.~Aker {\em et~al.}, (KATRIN Collaboration),
\newblock Nature Physics {\bf 18}, 160 (2022).

\bibitem{Aghanim2020}
{N. Aghanim} {\em et~al.}, (Planck Collaboration),
\newblock A\&A {\bf 641}, A6 (2020).

\bibitem{Agostini2020}
M.~Agostini {\em et~al.}, (GERDA Collaboration),
\newblock Phys. Rev. Lett. {\bf 125}, 252502 (2020).

\bibitem{Adgrall2021}
N.~Abgrall {\em et~al.}, (LEGEND Collaboration),
\newblock arXiv:2107.11462v1  (2021).

\bibitem{Adams2022}
D.~Q. Adams {\em et~al.}, (CUORE Collaboration),
\newblock Nature {\bf 604}, 53 (2022).

\bibitem{Alfonso2022}
K.~Alfonso {\em et~al.}, (CUPID Collaboration),
\newblock Journal of Low Temperature Physics  (2022), \\
https://link.springer.com/article/10.1007/s10909-022-02909-3

\bibitem{Anton2019}
G.~Anton {\em et~al.}, (EXO Collaboration),
\newblock Phys. Rev. Lett. {\bf 123}, 161802 (2019).

\bibitem{Albert2018}
J.~B. Albert {\em et~al.}, (nEXO Collaboration),
\newblock Phys. Rev. C {\bf 97}, 065503 (2018).

\bibitem{Dolinski2019}
M.~J. Dolinski, A.~W.~P. Poon, and W.~Rodejohann,
\newblock Annu. Rev. Nucl. Part. Sci. {\bf 69}, 219 (2019).

\bibitem{Eguchi2003}
K.~Eguchi {\em et~al.}, (KamLAND Collaboration),
\newblock Phys. Rev. Lett. {\bf 90}, 021802 (2003).

\bibitem{Abe2008}
S.~Abe {\em et~al.}, (KamLAND Collaboration),
\newblock Phys. Rev. Lett. {\bf 100}, 221803 (2008).

\bibitem{Araki2005b}
T.~Araki {\em et~al.}, (KamLAND Collaboration),
\newblock Nature {\bf 436}, 499 (2005).

\bibitem{Gando2011b}
A.~Gando {\em et~al.}, (KamLAND Collaboration),
\newblock Nature Geosci. {\bf 4}, 647 (2011).

\bibitem{Gando2012a}
A.~Gando {\em et~al.}, (KamLAND-Zen Collaboration),
\newblock Phys. Rev. C {\bf 85}, 045504 (2012).

\bibitem{Gando2013a}
A.~Gando {\em et~al.}, (KamLAND-Zen Collaboration),
\newblock Phys. Rev. Lett. {\bf 110}, 062502 (2013).

\bibitem{Shimizu2019}
I.~Shimizu and M.~Chen,
\newblock Front. Phys {\bf 7} (2019), \\
https://www.frontiersin.org/articles/10.3389/fphy.2019.00033

\bibitem{Gando2016}
A.~Gando {\em et~al.}, (KamLAND-Zen Collaboration),
\newblock Phys. Rev. Lett. {\bf 117}, 082503 (2016).

\bibitem{Abe2023}
S.~Abe {\em et~al.}, (KamLAND-Zen Collaboration),
\newblock Phys. Rev. Lett. {\bf 130}, 051801 (2023).

\bibitem{Gando2019}
A.~Gando {\em et~al.}, (KamLAND-Zen Collaboration),
\newblock Phys. Rev. Lett. {\bf 122}, 192501 (2019).

\bibitem{Kotila2012}
J.~Kotila and F.~Iachello,
\newblock Phys. Rev. C {\bf 85}, 034316 (2012).

\bibitem{Stoica2013}
S.~Stoica and M.~Mirea,
\newblock Phys. Rev. C {\bf 88}, 037303 (2013); updated in arXiv:1411.5506v3 .

\bibitem{PhysRevLett.111.142501}
N.~L. Vaquero, T.~R. Rodr\'{\i}guez, and J.~L. Egido,
\newblock Phys. Rev. Lett. {\bf 111}, 142501 (2013).

\bibitem{PhysRevC.91.024316}
J.~M. Yao, L.~S. Song, K.~Hagino, P.~Ring, and J.~Meng,
\newblock Phys. Rev. C {\bf 91}, 024316 (2015).

\bibitem{Rodriguez2010}
T.~R. Rodr{\'\i}guez and G.~Mart{\'\i}nez-Pinedo,
\newblock Phys. Rev. Lett. {\bf 105}, 252503 (2010).

\bibitem{Deppisch2020}
F.~F. Deppisch, L.~Graf, F.~Iachello, and J.~Kotila,
\newblock Phys. Rev. D {\bf 102}, 095016 (2020).

\bibitem{PhysRevC.91.034304}
J.~Barea, J.~Kotila, and F.~Iachello,
\newblock Phys. Rev. C {\bf 91}, 034304 (2015).

\bibitem{PhysRevC.101.044315}
L.~Coraggio, A.~Gargano, N.~Itaco, R.~Mancino, and F.~Nowacki,
\newblock Phys. Rev. C {\bf 101}, 044315 (2020).

\bibitem{Horoi2015}
A.~Neacsu and M.~Horoi,
\newblock Phys. Rev. C {\bf 91}, 024309 (2015).

\bibitem{Menendez2009}
J.~Menendez, A.~Poves, E.~Caurier, and F.~Nowacki,
\newblock Nucl. Phys. A {\bf 818}, 139 (2009).

\bibitem{PhysRevC.102.044303}
J.~Terasaki,
\newblock Phys. Rev. C {\bf 102}, 044303 (2020).

\bibitem{PhysRevC.91.024613}
J.~Hyv\"arinen and J.~Suhonen,
\newblock Phys. Rev. C {\bf 91}, 024613 (2015).

\bibitem{PhysRevC.87.045501}
F.~\ifmmode~\check{S}\else \v{S}\fi{}imkovic, V.~Rodin, A.~Faessler, and
  P.~Vogel,
\newblock Phys. Rev. C {\bf 87}, 045501 (2013).

\bibitem{PhysRevC.87.064302}
M.~T. Mustonen and J.~Engel,
\newblock Phys. Rev. C {\bf 87}, 064302 (2013).

\bibitem{PhysRevC.97.045503}
D.-L. Fang, A.~Faessler, and F.~\ifmmode~\check{S}\else \v{S}\fi{}imkovic,
\newblock Phys. Rev. C {\bf 97}, 045503 (2018).

\bibitem{DellOro2014}
S.~Dell'Oro, S.~Marcocci, and F.~Vissani,
\newblock Phys. Rev. D {\bf 90}, 033005 (2014).

\bibitem{Harigaya2012}
K.~Harigaya, M.~Ibe, and T.~T. Yanagida,
\newblock Phys. Rev. D {\bf 86}, 013002 (2012).

\bibitem{Asaka2020}
T.~Asaka, Y.~Heo, and T.~Yoshida,
\newblock Phys. Lett. B {\bf 811}, 135956 (2020).

\bibitem{Asai2020}
K.~Asai,
\newblock Eur. Phys. J. C {\bf 80}, 76 (2020).

\bibitem{Li2023}
A.~Li {\em et~al.},
\newblock Phys. Rev. C {\bf 107}, 014323 (2023).

\end{thebibliography}
\end{document}